\documentstyle[preprint,tighten,aps,psfig,epsf,floats]{revtex}

\newcommand{\ba}{\begin{eqnarray}}
\newcommand{\ea}{\end{eqnarray}}

\newcommand{\fr}[2]{\frac{#1}{#2}}
\newcommand{\non}{\nonumber}

\newcommand{\dr}{\partial_r}

\newcommand{\lb}{\left(}
\newcommand{\rb}{\right)}

\newcommand{\be}{\begin{equation}}
\newcommand{\ee}{\end{equation}}

\newcommand{\vrr}{{\bf r}}
\newcommand{\r}{{\bf r}}

\newcommand{\vp}{{\bf p}}
\newcommand{\vS}{{\bf S}}
\newcommand{\vL}{{\bf L}}

\newcommand{\sig}{\sigma}

\begin{document}

\preprint{TTP98-17, hep-ph/9805270}

\title{The $b$ quark low-scale running mass from 
$\Upsilon$ sum rules}

\author{Kirill Melnikov\thanks{
e-mail:  melnikov@particle.physik.uni-karlsruhe.de}}
\address{Institut f\"{u}r Theoretische Teilchenphysik,\\
Universit\"{a}t Karlsruhe,
D--76128 Karlsruhe, Germany}
\author{ Alexander Yelkhovsky\thanks{
e-mail:  yelkhovsky@inp.nsk.su}}
\address{ Budker Institute for Nuclear Physics\\
Novosibirsk, 630090, Russia}
\maketitle

\begin{abstract}
The $b$ quark low-scale running mass $m_{\rm kin}$
is determined from an analysis
of the $\Upsilon$ sum rules in the next-to-next-to-leading order (NNLO).
It is demonstrated that using this mass one can significantly
improve  the convergence of  the perturbation series for the
spectral density  moments. 
We obtain
$m_{\rm kin}(1{\rm GeV}) = 4.56 \pm 0.06~{\rm GeV}$.
Using this result we derive the value of the $\overline{MS}$
mass $\bar m$: $\bar m(\bar m) = 4.20 \pm 0.1~{\rm GeV}$.
Contrary to the low-scale running mass, the pole mass of the 
$b$ quark cannot be  reliably determined from the sum rules.
As a byproduct of our study
we find the NNLO analytical expression for the cross section 
$e^+e^- \to Q \bar Q$ of the quark antiquark pair production 
in the threshold region, as well
as the energy levels and the wave functions at the origin 
for the $^1S_3$ bound states of $Q\bar Q$.

\vspace{0.5cm}
\end{abstract} 

\section {Introduction}
The value of the bottom quark mass is an essential ingredient 
of the theoretical description of  $b$--hadrons. Among 
various applications, probably the most important one at present
is the determination of the Cabibbo-Kobayashi-Maskawa
matrix elements from $B$-decays.  Determination of the 
$b$ quark mass is based on the 
sum rules for the $\Upsilon$ mesons, proposed about 20 years 
ago \cite{Vol1,Vol2}. 
In the past several years, the sum rule analysis has been undertaken 
several times by different authors.
In particular, in the paper \cite{Vol} a 
very high accuracy of the $b$ quark pole mass was quoted.
The next attempt to extract the precise value of the $b$ quark pole  
mass from the sum rules was undertaken by Jamin and Pich in
\cite{JP}.  Their result differed from that of Ref. \cite{Vol}. 
The origin of the discrepancy
between these two results, as well as the flaws in both derivations, were
pointed out  by K\"uhn {\it et al.}~\cite{KPP}, 
who also determined the $b$ quark mass.
Two most recent papers on
the subject \cite{Penin,Hoangb} were devoted to an improvement
of the  theoretical accuracy of the mass determination and to a more
realistic estimate of the theoretical error.

In parallel to these developments, it became more and more clear
in the past years that the concept of the pole mass of a heavy 
quark is not a good one due to the intrinsic ambiguity 
of the order of $\Lambda_{\rm QCD}$ in 
its numerical value \cite{Bigi,BeBr}.
In contrast to this observation, all previous analyses were 
aimed at determining the pole mass of the $b$ quark from the 
sum rules. 
We note in this respect that {\em practical} problems 
in attempts to use the pole mass in heavy quark physics
are well appreciated; one of the vivid examples is provided
by the calculation of the inclusive semileptonic decay
widths of the $B$ mesons (see \cite{Klecture} for a review). 
In response to this problem,
it was pointed out how  the ``proper'' quark mass relevant 
for non--relativistic problems can be defined \cite{5infty}.  
In this paper we try to determine  this properly defined mass 
from the sum rules.

For the theoretical analysis
the moments of the photon polarization operator are used.
These moments can be computed analytically and compared to the experimental
ones. 
The moments of the photon vacuum polarization function
are defined through the dispersion 
integral:
\be
{\cal M}_n = \frac {12 \pi^2  M_1^{2n}}{n!} \frac {{\rm d}^n}{{\rm d}s^n}
\Pi (s) |_{s=0} = \frac {M_1^{2n}}{Q_b^2}
\int \limits_{0}^{\infty} \frac {R(s){\rm d}s}{s^{n+1}},
\label {moment}
\ee
where $R(s)$ is 
\be
R(s) = \frac {\sigma (e^+e^- \to b \bar b)}{\sigma_p},\qquad
\sigma _p = \frac {4\pi \alpha_{\rm QED}^2(m_b)}{3s}.
\ee
We defined the moments to be dimensionless by multiplying them
by the mass $M_1$ of the first $\Upsilon$ resonance in a 
suitable power.  Also, $Q_b=-1/3$ is the electric charge of the
$b$ quark in units of the positron charge.

The moments ${\cal M}_n$ can be calculated using experimental
input for $R(s)$. One gets: 
\be
{\cal M}_n^{\rm exp} = \frac {M_1^{2n}}{Q_b^2}
\left ( \frac {9\pi}{1.07\alpha_{\rm QED}^2} \sum \limits_{k}^{6}
\frac {\Gamma_k}{M_k^{2n+1}} + \int \limits_{s_0}^{\infty} 
\frac {R_c(s){\rm d}s}{s^{n+1}} \right ),
\label{momentexp}
\ee
where $M_k$ and $\Gamma_k$ are the masses and the electronic decay
widths of the first six $\Upsilon$ resonances and
$\alpha _{\rm QED}^2(m_b) = 1.07\alpha _{\rm QED}^2$ is used.
The $R_c(s)$ describes
the experimental spectral density, associated with the 
energy region above the open $B\bar B$ threshold; it is 
rather poorly known. The necessity to suppress
the contribution of this region is one of the reasons for 
using as high values of $n$ as possible. 

On the other hand, the same moments can be calculated theoretically,
using  the Operator Product Expansion (OPE) for the photon polarization
function.  The first non--perturbative
correction to the moments is associated 
with the gluon condensate \cite{Vol1,Vol2}.
It was shown that this contribution  
grows with $n$; however, for $n \le 20 $, the non-perturbative
contribution to the 
moments was estimated to be less than one per cent \cite{Vol2,Vol}.
Given exponential sensitivity of the moments to the value of the 
$b$ quark mass, the influence of the non--perturbative
corrections is clearly minor and can be neglected.
Therefore, the whole analysis for $n \sim 10$ reduces to 
a careful treatment of the perturbative effects in $R(s)$.
However, the perturbative treatment
is not simple, since 
for large values of $n$ the dominant
contribution to the perturbative moments 
comes from the threshold energy region. The relative
velocity of the $b \bar b$ system there 
is of the order of $\alpha _s$.
In this case, the 
theoretical spectral density can be calculated in the 
framework of the non--relativistic QCD, which means a simultaneous
expansion of the spectral density in $\alpha _s$ and in the relative 
velocity $\beta$ of the quark antiquark pair. As is well known,
the standard perturbation theory is not adequate in the threshold
region and the leading order approximation is the solution of the
Coulomb problem, which  resums all corrections 
of the form $(\alpha_s/\beta)^k$.
Going to NLO and NNLO, one calculates 
the spectral density $R(s)$ in the threshold region 
resumming all  
${\cal O}((\alpha _s/\beta)^k \times [1; \alpha_s, \beta;
\alpha_s^2,\alpha_s\beta,\beta^2])$ terms.
For this purpose we use the so called direct matching
procedure \cite{Hoang} which is described e.g. in \cite{HT,top,Hoangb}. 
We will not discuss all necessary details of this approach  
here and will merely quote the results of the
calculations. On the other hand, a part of the NNLO corrections
was treated numerically in \cite{HT,top}; for this reason, we present 
some additional theoretical results, which provide the imaginary
part of the polarization operator in the threshold region in 
completely analytical form to NNLO.

The theoretical expression for the spectral density
employed in this paper reads:
\be
R(s) = \lim_{r \to 0}{\mbox {Im}} \left [ N_c Q_b^2 \frac {24\pi}{s} 
\left (1-\frac {\vp ^2}{3m^2} \right )
G(\vrr,0) \right ],
\label {crsect}
\ee
where $G(\vrr,0)$ is the Green function of the non-relativistic 
Schr\"odinger equation
\be
(H-E-i\delta)G(\vrr,\vrr _1) = \delta ^{(3)}(\vrr -\vrr_1), 
\qquad E=\sqrt{s}-2m.
\ee
 Taken literally, Eq. (\ref{crsect}) is ill-defined due to 
improper treatment of the limit $r \to 0$ within the non--relativistic 
approach. In what follows, we circumvent this difficulty incorporating
the full QCD result and matching it with its non--relativistic 
counterpart. 

Integrating the theoretical expression for the spectral 
density in Eq.(\ref {moment}) and equating the obtained 
result  to the  experimental moments Eq.(\ref {momentexp})
one obtains the sum rules of the form:
\be
{\cal M}_n^{\rm exp} = {\cal M}_n^{\rm theor},
\ee
which give the numerical value for the $b$ quark mass, 
potentially with a small uncertainty.

To summarize, the usual
(and commonly used for the analysis) 
statement about theoretical calculations of the large-$n$ moments
can be formulated as follows: 1) for $n \sim 10-20$
the OPE ensures that non--perturbative corrections are small; 
2) for such $n$ the perturbative 
calculation of the moments requires the resummation of 
the Coulomb enhanced terms ${\cal O}(\alpha _s \sqrt{n})$.
This is achieved by calculating the spectral density 
$R(s)$ in the threshold region which requires a 
resummation of the  
${\cal O}((\alpha _s/\beta)^k) \times [1; \alpha_s, \beta;
\alpha_s^2,\alpha_s\beta,\beta^2])$ 
terms for the NNLO accuracy.   It is often assumed
implicitly, that this picture is correct independently of
all other parameters entering the sum rules
analysis. We do not believe that this is the case.
In particular, for the first point to be correct, 
it is crucial that one does not attempt to extract the {\em pole} mass 
of the heavy quark from the sum rules.

When this paper was prepared for publication, two 
other papers on this subject appeared \cite{Penin,Hoangb} 
where the NNLO analysis of the
$\Upsilon $ sum rules has been performed. The aim of both papers
was to extract the pole mass of the $b$ quark. In our approach to
the same problem we treat the so called
low energy running mass  as a quantity which can be determined
within the sum rule analysis; the pole mass of the $b$ quark 
is used at the intermediate stages of the calculation only. 

Let us present some heuristic arguments in favor of this approach.
It is known that the pole mass of the quark cannot be defined
when non--perturbative effects are addressed \cite{Bigi,BeBr}. 
It is also believed  that the bad behavior 
of the perturbation series in the relation,
say  between the pole mass and the $\overline {MS}$ mass, 
signals  this. The consequence of these 
facts is that there is  an irreducible ambiguity of the order 
of $\Lambda_{\rm QCD}$ in the numerical value of the pole mass.
If then the pole mass of the quark is used in the sum rules
analysis, its infrared sensitivity leads to  new 
infrared effects, which have no counterpart in the standard 
OPE \cite{vosem}. In particular, they are not described 
by the  gluon condensate \cite{Klecture}.

On the other hand, one can realize, that this is 
an artifact of the adopted procedure 
and the easiest way out of this problem is to abandon determination 
of the pole mass from the sum rule analysis.  
Therefore, we use the pole mass 
only as a tool to write the expression for the non--relativistic
Hamiltonian; however, we do not treat the pole mass 
as a fixed number and recalculate it 
consistently, order by order in perturbation theory.

Instead of the pole mass, one should determine some ``proper''
mass, which  does not suffer from 
a numerical ambiguity due to contributions of the soft momenta
region. Such proper masses are known -- one of the most familiar
in this respect is the $\overline {MS}$ mass. This will not be our
choice,  however. 

Elementary physical considerations suggest that the threshold 
problems are the low--scale problems, in principle. 
The typical scale is $\mu \sim m/\sqrt{n} \sim 1-2$ GeV 
for $n \sim 10$.
A  useful and reliable mass should therefore be normalized at such low
scale $\mu \ll m$. On the other hand, in order to 
suppress the contribution of the infra-red region, 
the inequality $\Lambda _{\rm QCD} \ll \mu$ should be respected.
The use of such low-scale running masses in various aspects
of heavy quark physics was repeatedly advocated in the last years
(for a detailed discussion and further references  see \cite{Klecture}).

The low-scale running mass cannot be defined uniquely,
because the only purpose of such definition is to remove the uncontrollable
contribution of the soft momenta region which affects
the pole mass. In this paper, we will work with 
the so called kinetic mass suggested in \cite{5infty}. 

The relation between the pole and the kinetic masses
is known to second order in $\alpha _s$
\cite{CzMeUr}:
\begin{equation}
m_{\rm pole} = m_{\rm kin}(\mu_Q) + \left [\Lambda (\mu_Q) \right ]_{\rm pert} +
\frac {1}{2 m_{\rm kin}(\mu_Q)} \left [ \mu^2_\pi(\mu_Q) \right ]_{\rm pert},
\label {polemass}
\end{equation}
where 
\begin{eqnarray}
\left [\Lambda (\mu) \right ]_{\rm pert} &=& 
\frac {4}{3}C_F\mu \frac {\alpha_s(\mu _1)}{\pi} \left \{
1 + \frac {\alpha_s}{\pi} \left [ 
\left (\frac {4}{3} - \frac {1}{2}\ln\frac{2\mu}{\mu_1} \right
)\beta_0 - C_A \left (\frac {\pi^2}{6} - \frac {13}{12} \right )
\right ] \right \},
\nonumber \\
\left [\mu_{\pi}^2 (\mu) \right ]_{\rm pert} &=& 
C_F\mu^2 \frac {\alpha_s(\mu _1)}{\pi} \left \{
1 + \frac {\alpha_s}{\pi} \left [ 
\left (\frac {13}{12} - \frac {1}{2} \ln \frac{2\mu}{\mu_1} \right
)\beta_0 - C_A \left (\frac {\pi^2}{6} - \frac {13}{12} \right )
\right ] \right \}.
\nonumber
\end{eqnarray}

For the rest of this paper, $\alpha _s$ denotes the strong coupling
constant in the $\overline {MS}$ scheme.
Also, we choose $\mu_Q$ to be equal to $1$ GeV, which 
seems to be a reasonable choice for the problem at hand. 
Then the ratio $\mu_Q/m$ is of the order of $\alpha _s$ 
and this gives the counting rule for the contributions
to the mass which should be accounted for when one goes from 
one order of perturbation theory to the other.
For example, to obtain the LO result we consider 
the pure 
Coulomb potential without any corrections
in the non--relativistic Hamiltonian for heavy quark antiquark
pair. Correspondingly,  the LO relation between the pole
and the kinetic masses is:
$$
m_{\rm pole} = m_{\rm kin}(\mu_Q)+\frac {4}{3} C_F \frac {\alpha_s}{\pi} \mu_Q.
$$
The NLO and the NNLO corrections to this
expression are added in accordance with the above counting rule.
Also, for our treatment, we use the same normalization scale 
$\mu_1$ for the strong coupling constant in the expression for the
mass,  as is used
in the non--relativistic Hamiltonian. This scale is called
$\mu_{\rm soft}$ in the rest of the paper.

We would like to note that, according to the above counting
rules, it would be necessary to know  
the term $\alpha _s ^3 \mu_Q$ in the 
relation between the pole mass and the kinetic mass. This term is
not known at present. The BLM-type estimate \cite {BLM} 
of this term is however
available and can be extracted from \cite{CzMeUr}.
Explicitly, the necessary term reads: 
\be\label{BLM}
\delta \left [\Lambda (\mu) \right ]_{\rm pert}  = 
\frac {4}{3}C_F\mu \left (\frac {\alpha_s(\mu _1)}{\pi} \right )^3
\left [\frac { \beta_0^2}{4} \left (\ln^2 \left (\frac { \mu}{\mu_1e} \right )
+1 \right )-
\left ( \frac {\beta_1}{8}+d_1\beta_0 \right )
\ln \left (\frac {\mu}{\mu_1e} \right )
+d_2 \right ],
\ee
where $\beta_0$ and $\beta_1$ are given in the next Section and 
$d_{1,2}$ read:
\ba
d_1 &=& \frac {\beta _0}{2} \left (\frac {5}{3} - \ln 2 \right )
 -C_A \left (\frac {\pi^2}{6} - \frac {13}{12} \right ),
\nonumber \\
d_2 &=&  \left (\frac {\beta _0}{2} \right )^2 \left [
\left (\frac {5}{3} - \ln 2 \right )^2 
-\left (\frac {\pi^2}{6}-\frac {31}{36} \right ) \right ].
\ea
In the numerical analysis of the last Section, we check the 
sensitivity of our results to the possible modification 
of the $d_2$ term due to additional terms which are not
accounted for in the BLM approximation. We find, that 
our final result for the kinetic mass is rather insensitive
to it.


As the result of our analysis, we find that  
the perturbation theory for the pole mass is not applicable:
typically, the NNLO corrections to the pole mass 
exceed the NLO ones and the dependence
of the result on the choice of the scale of the strong 
coupling constant is very strong.
These unwelcome features, therefore, do not permit  a reliable
determination of the $b$ quark  pole mass from the 
sum rules, with a trustworthy estimate of  the theoretical uncertainty.

On the contrary, the situation with the low-scale running mass 
looks more healthy: the perturbation series seem to be sign
alternating and the dependence on the normalization scale for 
the coupling 
constant appears to be reduced, as compared to the pole mass. 

The rest of the paper is organized as follows: in the next Section 
we discuss the framework of the calculation. In Section 3 the
corrections to the Green function due to corrections 
to the static quark antiquark potential are derived. In Section 4
the corrections to the Green function due to relativistic 
corrections to the heavy quark Hamiltonian are obtained. 
In Section 5 we combine these results and present the NNLO
expression for the theoretical spectral density $R(s)$ in the 
threshold region. In Section 6 the results for the 
energy levels and the wave functions at the origin 
for the $^1S_3$ $Q\bar Q$ resonances are  derived.
In Section 7 we present our final analysis 
for the sum rules and determine the low-scale mass 
of the $b$ quark. Finally we present our conclusions.

\section {The framework of the calculation}
We first discuss a framework of our calculations and introduce
all relevant notations. As we mentioned already, in order to obtain
the expression for the theoretical spectral density, we have to
calculate the expression for the imaginary part of the polarization
operator in the threshold region.
The threshold region is characterized
by a small value of the quark velocity $\beta$: 
\be
\beta =\sqrt{1-\frac {4m^2}{s}} \ll 1.
\ee
Here and below $m$ is the pole mass. The pole mass enters the 
usual perturbative expansion in quantum mechanics
for the nonrelativistic quarks. Later we will 
extract the low-scale running mass from the pole one.

Dynamics of  slow moving quark antiquark pair is governed
by the non--relativistic Hamiltonian:
\begin{eqnarray}
H &=& H_0 +V_1(r) + U(\vp,\vrr),
\nonumber \\
H_0 &=& \frac {\vp^2}{m} - \frac {C_Fa_s }{r},
\nonumber \\
V_1(r) &=& -\frac {C_F a_s^2}{4\pi r} \Bigg [
2\beta_0 \ln(2\mu'r) + a_1
\nonumber 
\\
&+&
\left (\frac {a_s}{4\pi} \right )
\left (\beta _0^2 \left (4\ln^2(\mu'r)+\frac {\pi^2}{3} \right )
+2(\beta _1 +2\beta_0a_1)\ln(\mu'r) + a_2 \right ) \Bigg ],
\nonumber \\
U(\vp,\vrr) &=& -\frac {\vp ^4}{4m^3}
 + \frac {\pi C_F a_s}{m^2} \delta ^3 (\vrr )
-\frac {C_F a_s}{2m^2r}
\left (\vp ^2 + \frac {\vrr ( \vrr  \vp ) \vp }{r^2} \right )
\nonumber \\
&+&\frac {3 C_F a_s}{2m^2r^3} \vS \vL
-\frac {C_F a_s }{2m^2} \left ( \frac {\vS^2}{r^3}-
3\frac {\left (\vS \vrr \right )^2}{r^5} - \frac {4\pi}{3} (2\vS^2 -3)
\delta(\vrr ) \right )-\frac {C_AC_F a_s^2}{2mr^2}.
\label {Ham}
\end{eqnarray}
In the above equations, the strong coupling constant is evaluated 
at the scale $\mu_{\rm soft}$:
\be
a_s = \alpha _s (\mu_{\rm soft}).
\ee
The scale $\mu '$ equals to 
$\mu e^{\gamma _E}$.

The operator $U(\vp, \vrr)$ is the QCD generalization of the 
standard Breit potential \cite {LL}. The last term in the expression 
for  the operator $U(\vp, \vrr)$
is the non--Abelian contribution, originating from a
correction to the Coulomb exchange, caused  by a 
transverse gluon \cite{NABreit}.
The potential $V_1(r)$ represents a deviation of the 
static QCD potential from the Coulomb one. It was calculated
to order $\alpha _s ^2$ in \cite{Fishler} and to order
$\alpha _s ^3$ in \cite {Peter}.  The coefficients there
read explicitly:
\begin{eqnarray}
\beta _0 &=& \frac {11}{3} C_A- \fr{4}{3} N_LT_R,
\nonumber \\
\beta _1 &=& \frac {34}{3} C_A^2 - \frac {20}{3} C_AT_RN_L - 4C_F T_R N_L,
\nonumber \\
a_1 &=& \frac {31}{9} C_A - \frac {20}{9} T_R N_L ,
\nonumber \\
a_2 &=& \left (\frac {4343}{162} + 6\pi^2 - \frac {\pi^4}{4}+\frac
{22}{3} \zeta_3 \right )C_A^2 -
\nonumber \\
&& 
\left (\frac {1798}{81}+\frac {56}{3}\zeta_3
\right ) C_A T_R N_L
-\left (\frac {55}{3} - 16\zeta_3 \right )C_FT_RN_L + \left (\frac
{20}{9} T_R N_L \right )^2.
\end{eqnarray}

The $SU(3)$ color factors are
$C_A = 3, C_F = 4/3, T_R = 1/2$. $N_L = 4$ is the number of quarks
whose masses have been neglected.

Given the Hamiltonian $H$, one can find the  Green function
of the Schr\"odinger equation:
\be
(H-E-i\delta)G(\vrr,\vrr _1) = \delta ^{(3)}(\vrr -\vrr_1), 
\qquad E=\sqrt{s}-2m.
\ee
Once the Green function is found, the non--relativistic 
cross section for the $Q\bar Q$ pair production in $e^+e^-$
annihilation is obtained using Eq.(\ref {crsect}).

Treating the corrections to the Green function in the perturbation
theory, one can consider the corrections due to $V_1(r)$ and 
$U(\vp,\vrr)$ separately.  The corrections to the Green function 
due to $U(\vp,\vrr)$ were recently calculated in \cite{HT,top}.
These corrections are not simple conceptually, because
they deliver divergent contributions to the Green function at
the origin. The divergences are removed by matching the
result of the calculations in quantum mechanics to the
result of the full QCD calculation \cite {CzM}.
 Technically, however, the calculation of the correction
caused by $U(\vp,\vrr)$ is very simple 
and can be performed algebraically 
(see \cite{top}). On the other
hand, the corrections to the imaginary part of the Green function 
due to the $V_1(r)$ perturbation can be calculated within the 
quantum mechanics and for this reason these corrections  
are rather simple conceptually. However, they provide
the most challenging part of the whole calculation from the technical 
viewpoint. For this reason, the calculation of the 
corrections to the imaginary part of the Green function due 
to the $V_1(r)$ perturbation is discussed below in some detail.

\section{Corrections to the Green function due to $V_1(r)$}

\subsection{The Coulomb Green function}

In this section we collect useful  formulas for the
Coulomb Green function. 
A convenient expression for the $S$-wave Coulomb 
Green function can be found in \cite{MS}: 
\be
G(r,r_1) = \frac {-ime^{ik(r+r_1)}}{4\pi\sqrt{rr_1}}\int \limits_{0}^{\infty}
\frac {{\rm d}t}{\sqrt{t(t+1)}} \left (\frac {1+t}{t} \right )^{i\nu}
e^{2ik(r+r_1)t}J_1(4k\sqrt{rr_1}\sqrt{t(t+1)} ),
\label {Grr1}
\ee
In the above expression,  
$\nu = C_F\alpha_s/(2\beta)$ and $k=m\beta$.

Using this representation, one easily obtains the expression for 
$G(r,0)$:
\be
G(r,0) = \frac {-imk}{2\pi}e^{ikr}\int \limits _{0}^{\infty}
{\rm d}t \left (\frac {1+t}{t} \right )^{i\nu} e^{2ikrt}.
\label {Gr0}
\ee

 From the expression for the cross section (cf. Eq. (\ref {crsect})), 
it is clear,  that  one is interested in the behavior of the 
Coulomb  Green function for small $r$.
For small $r$, the Green function diverges like $1/r$; the principal
divergence is related to the behavior of the free ($\alpha_s=0$)
Green function:
\be
G^{(0)}(r,0) = \frac {m}{4\pi r}e^{ikr}.
\label {Gfree}
\ee

\subsection{Generating function}
To calculate  corrections to the Coulomb Green function 
at the origin caused
by the $V_1(r)$ perturbation, 
it is convenient to introduce a generating function $g(\sigma)$:
\be\label{g}
g(\sigma)= \int d^3\vrr\; G^2(r,0)\fr{(2\mu r)^{\sig}}{r}.
\ee
Once this function is found, one easily obtains a correction 
to the Green function at the origin caused by the 
$V_1(r)$ term in the potential.
For further calculations, it will be convenient to separate the
generating function into two terms:
\be
g(\sigma) = g_{\rm free}(\sigma) + g_1(\sigma),
\label {gsigma}
\ee
where
\begin{eqnarray}
g_{\rm free}(\sigma) &=&  
\int {\rm d}^3\vrr\; G_0^2(r,0)\fr{(2\mu r)^{\sig}}{r},
\nonumber \\
g_1(\sigma) &=&  \int {\rm d}^3\r \;
\left \{G^2(r,0) -  G_0^2(r,0) \right \}
\fr{(2\mu r)^{\sig}}{r}.
\label {g1}
\end{eqnarray}
In the above equations, $G_0(r,0)$ is the free Green function
(cf. Eq. (\ref {Gfree})).

It is an easy task to calculate $g_{\rm free}$. One gets:
\be
g_{\rm free} = \frac{m^2}{4\pi} \left (\frac {i\mu}{k}\right )^{\sigma}
\Gamma (\sigma).
\label {gsigfree}
\ee

We need further $g_1(\sigma)$.
For the
Coulomb Green function we  use the representation from Eq. (\ref {Gr0}).
Integrating then over $\vrr$ in Eq. (\ref{g1}), one gets:
\be
g_1(\sigma)=\fr{m^2}{4\pi} \lb \fr{ i\mu }{ k } \rb^{\sigma}
          \Gamma(2+\sigma)
          \int \limits_0^{\infty}dtds \left [ \lb \fr{1+t}{t} \rb^{i\nu}
          \lb \fr{1+s}{s} \rb^{i\nu} -1 \right ]
          (1+t+s)^{-2-\sigma}.
\ee
To proceed further, it is convenient to introduce new integration variables,
\be
\tau=\fr{t}{1+t},\qquad \rho=\fr{s}{1+s}.
\ee
Then one gets:
\be
g_1(\sigma)=\fr{m^2}{4\pi} \lb \fr{ i\mu }{ k } \rb^{\sigma}
          \Gamma(2+\sigma)
          \int \limits_0^{1}d\tau d\rho \fr{\tau^{-i\nu} \rho^{-i\nu} -1}
          {(1-\tau)^{-\sigma}
          (1-\rho)^{-\sigma}}
          (1-\rho\tau)^{-2-\sigma}.
\ee

Substituting also 
$\eta=\rho\tau$ we arrive at:  
\be
g_1(\sigma)=\fr{m^2}{4\pi} \lb \fr{ i\mu }{ k } \rb^{\sigma}
          \Gamma(2+\sigma)
          \int \limits_0^{1}d\eta   
           \fr{ \eta^{-i\nu} -1 }{ (1-\eta)^{2+\sigma} }
          \int \limits_{\eta}^{1}\fr{d\tau}{\tau}(1-\tau)^{\sigma}
          \lb 1-\fr{\eta}{\tau} \rb^{\sigma}.
\ee
Finally, changing the variables $\tau \to \xi$ with
$\xi=(1-\tau)/(1-\eta)$, we get
\be
g_1(\sigma)=\fr{m^2}{4\pi} \lb \fr{ i\mu }{k} \rb^{\sigma}
          \Gamma(2+\sigma)
\int \limits_0^{1}d\eta \fr{ \eta^{-i\nu}-1 }{ (1-\eta)^{1-\sigma} }
\int \limits_0^{1}d\xi \xi^{\sigma}(1-\xi)^{\sigma}\lb 1-(1-\eta)\xi
\rb^{-\sigma-1}.
\ee
The last integral is proportional to
$F_{21}(1+\sigma,1+\sigma;2+2\sigma;1-\eta)$ and equals to
\be
\int \limits_0^{1}{\rm d}\xi \fr{\xi^{\sigma}(1-\xi)^{\sigma}}
{\lb 1-(1-\eta)\xi \rb^{\sigma+1}} =
\sum_{n=0}^{\infty}\fr{(1+\sigma)_n^2}{(n!)^2}
\left[2\psi(n+1)-2\psi(1+n+\sigma)-\ln\eta\right]\eta^n,
\ee
where the series representation for the hypergeometric function was
used. 
Here $(z)_n=\Gamma(z+n)/\Gamma(z)$ is the Pochhammer symbol.
Integrating over $\eta$, we
obtain
\ba
g_{1}(\sigma)=\fr{m^2}{4\pi}
          \lb \fr{ i\mu }{k} \rb^{\sigma}
          \Gamma(2+\sigma) \Gamma(\sigma)
\sum_{n=0}^{\infty}\fr{(1+\sigma)_n^2}{(n!)^2}
\left[2\psi(n+1)-2\psi(n+1+\sigma)-\partial_n \right]\times&& \non \\
\left\{
        \fr{ \Gamma(n+1-i\nu) }{ \Gamma(n+1-i\nu+\sigma) }
      - \fr{ \Gamma(n+1) }{ \Gamma(n+1+\sigma) }           \right\}&&,
\label{g1fina}
\ea
which can be presented in a more compact form\footnote{We thank O. Yakovlev
for pointing this to us.}:
\be
g_{1}(\sigma)=-\fr{m^2}{4\pi}
                \lb \fr{ i\mu }{ k } \rb^{\sigma}
                \fr{ 1+\sigma }{ \sigma }
                \sum_{n=1}^{\infty} \partial_n
                \left\{
                \fr{ \Gamma^2(n+\sigma) }{ \Gamma^2(n) }
                \fr{ \Gamma(n-i\nu) }{ \Gamma(n-i\nu+\sigma) }
                - \fr{ \Gamma(n+\sigma) }{ \Gamma(n) } \right\} .
\label {g1fin}
\ee
In Eqs.(\ref{g1fina},\ref{g1fin}) $\partial _n$ stands for the
partial derivative with respect to $n$.
Eqs. (\ref{gsigma},\ref{gsigfree},\ref{g1fin}) 
provide a necessary expression for the generating function.

\subsection{Correction to the Green function 
due to $\ln^{n}(2\mu r)/r,\; n \le 2$ perturbation}

To evaluate these corrections we will use the representation
of the generating function $g(\sigma)$ given in the previous
section. For the case of interest, the correction to the imaginary 
part of the Green function at the origin due to $\ln^{n}(2\mu r)/r$
perturbation is obtained as the $n$-th derivative of 
the imaginary part of the generating function with respect to 
$\sigma$ at $\sigma \to 0$. To obtain these derivatives, we first expand
the generating function up to the second order in $\sig$:
\ba
g_{\rm free}(\sigma) &=& \fr{m^2}{4\pi}
\left\{ \fr{1}{\sigma} + \ln \fr{ i\mu }{ k } - \gamma_E +
\fr{\sigma}{2} \left[ \lb \ln \fr{ i\mu }{ k } - \gamma_E \rb^2 +\fr {\pi^2}{6}
\right] \right. \non \\
&& \left.
    + \fr{\sigma^2}{6} \left[ \lb \ln \fr{ i\mu }{ k } - \gamma_E \rb^3
    + \fr {\pi^2}{2} \lb \ln \fr{ i\mu }{ k } - \gamma_E \rb -2\zeta _3
      \right] + \ldots \right\};
\ea
and
\be
g_{1}(\sigma)=\frac {m^2}{4\pi} \left ( g_{1}(0)+
\sigma g'_{1}(0)
+ \fr{\sigma^2}{2} g''_{1}(0) + \ldots, \right )
\ee
where
\ba
g_{1}(0)&=& - 
         \sum_{n=1}^{\infty} \partial_n \lb \psi(n) - \psi(n-i\nu) \rb
         = -\gamma_E - \partial_{i\nu} i\nu \; \psi(1-i\nu), \\
g'_{1}(0)&=& \lb \ln \fr{ i\mu }{ k } + 1 \rb
         g_{1}(0) +\fr{\pi^2}{6} 
         +\fr{1}{2}\partial_{i\nu}^2 i\nu \; \psi(1-i\nu)\non \\
         &&
         - \fr{1}{2}\sum_{n=1}^{\infty} \partial_n
         \left[ 
         \lb \psi(n) - \psi(n-i\nu) \rb \lb 3\psi(n) - \psi(n-i\nu) \rb
         \right], \\
g''_{1}(0)&=& 2 \lb \ln \fr{ i\mu }{ k } + 1 \rb
         g'_{1}(0)
         - \left \{ \lb \ln \fr{ i\mu }{ k } + 1 \rb^2 +1 \right \}
            g_{1}(0) 
            -\fr{1}{3}\partial_{i\nu}^3 i\nu \; \psi(1-i\nu)\non \\
         &&- \fr{1}{3}\sum_{n=1}^{\infty} \partial_n
         \left\{ 
         \fr{3}{2} \lb 3 \psi^2(n)-4\psi(n)\psi(n-i\nu)+\psi^2(n-i\nu) \rb'
         \right. \non \\
         && \left.
         +7\psi^3(n)-12\psi^2(n)\psi(n-i\nu)+6 \psi(n)\psi^2 (n-i\nu)
         -\psi^3(n-i\nu) \right\}.
\ea

Using these expressions, we easily find the corrections
to the imaginary part of the Green function at the origin,
caused by  $\ln^n(2\mu r)/r$ perturbations. In the formulas below
we disregard the $1/\sigma$ pole which is present in 
the expression for $g_{\rm free}$
since it does not contribute to the imaginary part.
We obtain explicitly:
\be
\delta G_L(\mu) = -\int {\rm d}^3 \vrr G(r,0)^2 \frac {\ln(2 \mu r)}{r} =
\frac {-m^2}{4\pi} \left \{\frac {1}{2} 
\left[ \lb \ln \fr{ i\mu }{ k } - \gamma_E \rb^2 +\fr {\pi^2}{6} \right
]
+g_1'(0) \right \}.
\ee

In a similar manner one obtains the correction due to $\ln^2(2\mu r)/r$
perturbation:
\ba
\delta G_{L2}(\mu)&=&-\int {\rm d}^3 \vrr G(r,0)^2 \frac {\ln^2(2 \mu r)}{r} =
\nonumber \\
&&
\frac {-m^2}{4\pi} \left \{ \frac {1}{3}
 \left[ \lb \ln \fr{ i\mu }{ k } - \gamma_E \rb^3
    + \fr {\pi^2}{2} \lb \ln \fr{ i\mu }{ k } - \gamma_E \rb -2\zeta _3
      \right] + g''_1(0) \right \}.
\ea

\subsection{The second iteration of the logarithmic perturbation}

The leading term of 
the static potential $V_1(r)$ provides an ${\cal O}(\alpha _s)$
correction. Therefore, one should calculate the second order
correction induced by the perturbation $\ln(2\mu r)/r$.
Such calculation is described below.

Explicitly, we have  to calculate:
\be
\delta G^{(2)}(\mu) = \int {\rm d}^3 \vrr\;{\rm d}^3 \vrr_1 G(r,0)V(r)
G(r,r_1)
V(r_1)G(r_1,0),~~~~V(r) = \ln(2\mu r)/r.
\label{secit}
\ee
For this purpose we use the representation for the 
Green function $G(r,r_1)$, given in Eq. (\ref {Grr1}).
We  define:
\be
G_1(\sig,\sig _1) = \left (\frac {i\mu}{k} \right )^{\sig +\sig_1} 
\frac {1}{\Gamma(-\sig)\Gamma(-\sig_1)}
\int \limits_{0}^{\infty} \frac {{\rm d}\tau {\rm d}\tau _1}{\tau ^{1+\sig}
\tau_1^{1+\sig_1}} F(\tau, \tau_1),
\ee
where
\be
F(\tau,\tau_1) = \int {\rm d}^3\r {\rm d}^3\r_1G(r,0)
\frac {e^{2ikr\tau}}{r}G(r,r_1)
\frac {e^{2ikr_1\tau_1}}{r_1} G(r_1,0).
\ee
Then
\be
\delta G^{(2)} = 
\frac {\partial^2}{\partial _\sig \partial _{\sig _1}} 
G_1(\sig,\sig_1)|_{\sig=0,\sig_1=0}.
\label{G2}
\ee

Using Eqs. (\ref {Grr1},\ref {Gr0}) 
for the Green functions, 
and performing the integrations over $r$, $r_1$ and the variable
$t$ which enters the integral representation of the $G(r,r_1)$
function, we arrive at the following 
result:
\be
F(\tau,\tau_1) = \frac {im^3}{8\pi k (1-i\nu)} F_1(\tau,\tau_1),
\ee
\be
F_1(\tau,\tau_1) = 
\int \limits_{0}^{1}  \frac {{\rm d}u\;{\rm d}v}{\Delta^2 \Delta_1^2}
\left (\frac {1+u}{u} \right )^{i\nu} \left (\frac {1+v}{v} \right )^{i\nu}
F_{21} \left (2,1-i\nu;2-i\nu;\frac {(\Delta-1)(\Delta_1-1)}
{\Delta \Delta_1} \right ),
\ee
where
\be
\Delta = 1+u+\tau,\qquad \Delta_1 = 1+v+\tau_1,
\ee
and $F_{21}(a,b;c;z)$ is the Gauss hypergeometric function.

The function $G_1(\sig,\sig_1)$ can be written as
\be
G_1(\sig,\sig_1) =  \left (\frac {i\mu}{k} \right )^{\sig +\sig_1}
\frac {im^3}{8\pi k(1-i\nu)}W(\sig,\sig_1),
\ee
where
\be
W(\sig,\sig_1) = \frac {1}{\Gamma(-\sig)\Gamma(-\sig_1)}
\int \limits_{0}^{\infty} \frac {d\tau\;d\tau_1}{\tau^{1+\sig}\tau^{1+\sig_1}}
F_1(\tau,\tau_1).
\ee

For further calculation, it is convenient to 
define new variables
$$
u \to u_1 = u -\tau;\qquad v \to v_1 = v-\tau_1
$$
and integrate over $\tau $ and $\tau _1$.
Performing another variable transformation,
$$
u_1 \to x =\frac {u_1}{1+u_1}\qquad v_1 \to y = \frac {v_1}{1+v_1},
$$
we finally get:
\begin{eqnarray}
W(\sig,\sig_1)&=&\frac {\Gamma(1-i\nu)^2}{\Gamma(1-i\nu-\sig) 
\Gamma(1-i\nu-\sig_1)} \int \limits_{0}^{1} dx\;dy
\left (\frac {x}{1-x} \right )^{-\sig} 
\left (\frac {y}{1-y} \right )^{-\sig _1}
(xy)^{-i\nu} 
\nonumber \\
&\;&
F_{21}(-i\nu,-\sig;1-i\nu-\sig;x) 
F_{21}(-i\nu,-\sig _1;1-i\nu-\sig _1;y)
F_{21}(2,1-i\nu;2-i\nu;xy).
\nonumber 
\end{eqnarray}

One notes that if the last hypergeometric function in the above
equation is expanded  
in Taylor series in $xy$,  integrations over $x$ and $y$
factorize. These series would, however, diverge for $xy=1$.
Nevertheless, upon integration over $x$ and $y$, one gets a
 series
which converges as $1/n^2$. 
We conclude therefore,  that this operation is legitimate.
We write  $W(\sig,\sig_1)$ in a factorized form:
\be
W(\sig,\sig_1) =
\frac {\Gamma(1-i\nu)^2 (1-i\nu)}{\Gamma(1-i\nu-\sig)
 \Gamma(1-i\nu-\sig_1)}
\sum \limits_{n=1}^{\infty} \frac {n}
{n-i\nu} T(n,\sig)T(n,\sig_1),
\ee
where
\be
T(n,\sig) = \int \limits_{0}^{1} dx x^{n-1-i\nu}
\left (\frac {x}{x-1} \right )^{-\sig} F_{21}(-i\nu,-\sig;1-i\nu -\sig;x).
\label {Tn}
\ee

According to Eq.(\ref {G2}), one needs an expansion
of  $T(n,\sig)$ in  $\sig$ up to the first power.
This can be easily done by expanding the 
hypergeometric function in Taylor series in Eq.(\ref {Tn}), 
evaluating resulting integrals, extracting the limit $\sig \to 0$ 
and then resumming the resulting series. 
We obtain:
\be
T(n,\sig) = \frac {1}{n-i\nu} -\sig T_1(n),
\ee
where
\be
T_1(n) = \frac {2\gamma}{n-i\nu} + \frac {\psi(1-i\nu)}{n}
+\frac {\psi(n-i\nu)(n+i\nu)}{n (n-i\nu)}+
\frac {i\nu}{n(n-i\nu)^2}.
\ee
With this result, we obtain the series representation 
for the function $W(\sig ,\sig_1)$ with the required 
accuracy in $\sig,\sig_1$.
Upon differentiation over $\sig$ and $\sig_1$ at $\sig = \sig _1 = 0$
we get the correction to the Green function $\delta G^{(2)}$:
$$
\delta G^{(2)}(\mu) = \frac {im^3}{8\pi k}\left \{
\sum \limits_{n=1}^{\infty} \frac {n}{n-i\nu} \left (
\frac {\ln\left (\frac {i\mu}{k} \right )+\psi(1-i\nu)}{n-i\nu}
-T_1(n) \right )^2 \right \}.
$$

\section{Corrections to the Green function due to the $U(\vp,\vrr)$
perturbation}

We now briefly discuss  corrections to the Green function at the origin
caused by the operator $U(\vp,\vrr)$ from Eq.(\ref {Ham}). 
This correction is obtained as
\be
\delta G^{U} = -\int {\rm d}^3 \vrr G(r,0)\;U(\vp,\vrr)\; G(r,0).
\label {pert}
\ee
As long as we are interested in the $Q\bar Q$ pairs, produced
in the triplet $S$--states, only the corresponding 
projection of the operator $U(\vp,\vrr)$ should be considered.
Substituting $\vS^2 =1 $ and ${\bf S}{\bf L}=0$ in Eq.(\ref {Ham}),
it is easy to get that $U(\vp,\vrr)$ can be presented in the following form:
\be
U(\vp,\vrr) = 
-\frac {\vp ^4}{4m^3} + \frac {11\pi a_s C_F}{3m^2} \delta ^{(3)} (\vrr )
-\frac {C_F a_s}{2 m^2} \left \{ \frac {1}{r},\vp ^2 \right \}
-\frac {C_AC_Fa_s^2}{2mr^2}.
\label {Br}
\ee

At this stage, it is advantageous to express this operator in terms of
the zeroth order Hamiltonian $H_0$ in order to apply the equation of
motion for the Green function $G(r,0)$:
\be
(H_0-E)G(r,0) = \delta^{(3)}(\vrr).
\label {eqmot}
\ee

This is most easily done  using the following commutation relations:
\ba
 \left [ H_0, ip_r \right ] &=& \fr{4\pi\delta^{(3)}(\vrr)}{m}
 +\fr{2\vL ^2}{mr^3}-
 \fr{C_Fa_s}{r^2},
\\
\left\{H_0,\fr{1}{r}\right\}&=&\fr{2}{r}H+
\fr{4\pi\delta^{(3)}(\vrr)}{m}+\fr{2}{mr^2}\dr,
\ea
where $p_r = -i (\dr + 1/r )$ is the radial momentum operator. 
One finds that the operator $U(\vp,\vrr)$
from Eq. (\ref{Br}) can be written as:
\be
\label{Umod}
U(\vp,\vrr) = - \fr{ H_0^2 }{4m}
              - \fr{ 3C_Fa_s }{4m} \left\{H_0,\fr{1}{r} \right\}
              + \fr{ 11 C_F a_s }{12m} \left [ H_0, ip_r \right ]
              - \fr{ \lb 2C_F + 3C_A \rb C_Fa_s^2 }{ 6mr^2 }.
\ee

Let us consider the first three terms of Eq. (\ref {Umod}).
Inserting them into Eq. (\ref {pert}) and using the equation of motion
for the Green function (\ref {eqmot}),  we find:
\ba
&&- \int {\rm d}^3 \vrr G(r',r)
  \lb -\fr {H_0^2}{4m}- \fr{ 3C_Fa_s }{4m} \left\{H,\fr{1}{r} \right\}
              + \fr{ 11 C_Fa_s }{12m} [ H, ip_r] \rb G(r,r'') \non \\
&& = \left [ \fr {E}{2m} + \fr{ 3C_Fa_s }{2mr} \right ] G(r,0)
  +2ip_r G(r,0)
   + \int {\rm d}^3 \vrr G(0,r) \left \{ \fr {E^2}{4m} +
\fr{ 3C_Fa_s E }{2mr} \right \} G(r,0).
   \label{cc}
\ea
All terms in the above equation which cannot
contribute to the imaginary part of the Green function have been
omitted.

Also, it is easy to recognize that the terms in Eq. (\ref {cc}), which 
still have to be integrated  over $\vrr$, can be easily obtained if one 
redefines the eigenvalue and the coupling constant of the lowest
order Hamiltonian $H_0$:
\be
H_0 \to {\cal H} =  \frac {\vp^2}{m} - \frac {C_Fa_s}{r}\left (1+ \frac
{3E}{2m} \right);\qquad E \to {\cal E} = E+\frac {E^2}{4m} = \frac
{p_0^2}{m} = \frac {m\beta^2}{1-\beta^2}.
\ee
The new Hamiltonian ${\cal H}$ is still of the Coulomb form and,
therefore, the solution for the Coulomb Green function
presented in the previous section can be used.

Therefore, we conclude, that the  
only non--trivial calculation required here is  
 the correction to the Green function at the origin caused by 
the $1/r^2$ perturbation, which is explicitly given by the last term
in Eq. (\ref {Umod}). We refer the reader to the paper \cite{top}
where the 
details of the calculation are discussed. 

\section {Complete result and matching}

We now combine the results of the above calculations 
and write the final result in the form:
\begin{eqnarray}
R(s) &=& {\cal K}(\mu_{\rm hard},\mu_{\rm fact})\; ( R_1(s) + R_2(s)),
\label {rs}
\\
R_1(s) &=& \frac {3}{2} N_c Q_b^2 C_F a_s
\mbox{Im} \left \{ H(C_F a_s,\beta)  
\left [ 1 + C_F a_s^2 \left (\frac { C_F}{3}
+ \fr{C_A}{2}  \right ) H(C_F a_s,\beta )\right ]
\right \},
\nonumber \\
R_2(s) &=& \frac {6\pi C_F}{m^2} N_c Q_b^2 \mbox{Im} \Bigg \{ 
\frac {-2 a_s^2}{4\pi} 
\left [\beta _0 + \frac {a_s}{4\pi}
\left ( \beta _1 + 2\beta _0 a_1 \right )
\right ]
\delta G_L\left (\frac {\mu_a}{2} \right )
-
\frac {4\beta _0^2 a_s^3}{(4\pi)^2}
\delta G_{L2}\left (\frac {\mu_1}{2} \right )+
\nonumber \\
&&\frac {4\beta _0^2C_F a_s^4}{(4\pi)^2} 
\delta G^{(2)}\left (\frac {\mu_b}{2} \right )
\Bigg \}.
\nonumber \\
\end{eqnarray}
In the above expression, $R_2$ is the contribution due to the
$V_1(r)$ perturbation. The scales $\mu_1,\mu_a,\mu_b$ there read
explicitly:
\be
\mu_1 = \mu_{\rm soft} \exp[\gamma_E],\qquad
\mu_a = \mu_1 \exp\left [\frac {a_1+\frac {a_s}{4\pi} (\frac
{\pi^2}{3} \beta_0^2 + a_2)}{2\beta_0 + 
\frac {a_s}{4\pi} 2 (\beta _1 + 2\beta_0a_1 )} \right ],\qquad
\mu _b = \mu_1 \exp \left [ \frac {a_1}{2\beta _0} \right ].
\label{scales}
\ee

The function $H(a,\beta)$ was first obtained
in a similar context in \cite {Hoang} and is given by:
\be
H(a,\beta) = \left (1-\frac {\beta ^2}{3} \right )
\left \{\frac {i\beta}{a} - (1+\beta ^2) \left [
\gamma _E + \ln \left (\frac {-i\beta m}{\mu_{\rm fact}} \right )
+\psi \left (1-ia \frac {1+\beta ^2}{2\beta} \right ) \right
] \right \}.
\ee

We have absorbed all energy-independent divergent contributions 
to the factor ${\cal K}(\mu_{\rm hard},\mu_{\rm fact})$, which is determined
by matching  the above result for $R(s)$ to
the result of the perturbative calculations in full QCD  
\cite {CzM} in the region 
$\alpha _s \ll \beta \ll 1$, where both results are supposed to be
valid. One gets:
\ba
{\cal K}(\mu_{\rm hard},\mu_{\rm fact})&=& 1 + 
C_1 C_F \left (\frac {\alpha _s(\mu_{\rm hard})}{\pi} \right ) + 
C_2 C_F \left (\frac {\alpha _s(\mu_{\rm hard})}{\pi} \right )^2,
\label {hard}
\\
C_1 = -4; \qquad && C_2 = C_F C_2^{A} + C_A C_2^{NA}+T_R N_L C_2^{L} + T_H
N_H C_2^{H}-C_1\frac{\beta_0}{4} \ln \left (\frac {m^2}{\mu_{\rm hard}^2}
\right ),
\ea
and
\begin{eqnarray}
C_2^{A}&=&\frac {39}{4} -\zeta _3 +\pi ^2 \left (\frac
{4}{3}\ln2-\frac {35}{18} \right )+\frac {\pi ^2}{3} \ln \frac
{m^2}{\mu_{\rm fact}^2} ;
\nonumber \\
C_2^{NA}&=& -\frac {151}{36}-\frac{13}{2} \zeta _3 
+\pi^2 \left (\frac {179}{72} -\frac {8}{3}\ln2 \right )
+\frac {\pi^2}{2}\ln \frac {m^2}{\mu_{\rm fact}^2} ;
\nonumber \\
C_2^{H} &=&\frac {44}{9} - \frac {4}{9}\pi^2;
\nonumber \\
C_2^{L} &=& \frac {11}{9}.
\end{eqnarray}

\section{Corrections to the energy levels and the wave functions}

It is known, that the proper expression for the Green function 
which is valid in the whole threshold energy region is given 
by the expression:
\be
G(E+i\epsilon;0,0)
=
      \sum_n \fr{ |\Psi_n|^2  }{ E_n - E - i\epsilon}
              + \int \limits_0^{\infty} \fr{ dk }{ 2\pi }
                     \fr{ |\Psi_k|^2 
                     }{ E_k - E - i\epsilon }.
\label {rGF}
\ee
Here $E_n$ and $\Psi_n$ are the energy levels and the wave functions
at the origin of the perturbative $^3S_1$ $\bar Q Q$ resonances, 
which can be calculated order by order in perturbation theory. 
On the other hand, when a correction
to the Green function is calculated as a power series over a 
perturbation, the 
energy denominators entering the exact expression (\ref{rGF}) are 
also expanded in powers of $\alpha _s$. 
It is therefore possible to extract the corrections to the energy 
levels and to the wave functions of the resonances by performing 
the Laurent expansion around $E=E_n^{(0)}$ of the 
corrections to the Green function obtained in the previous sections.
On one hand, these results are interesting by itself, 
for applications to various problems that involve perturbative calculations
for bound states both in QED and QCD. On the other hand, 
they are used below to construct the proper theoretical expression 
for the large--$n$ moments.

In the formulas below we denote:
$$
\psi_z = \psi(z),\; \psi_z'=\frac {{\rm d}}{{\rm d}z} \psi(z),\;{\rm etc.}
$$

To present our results, it turns out to be useful to 
define a function
\be
S_i(n) = \sum \limits_{k=1}^{n-1} \frac {\psi _k}{k^i}.
\ee

The energy levels and the wave functions at the lowest
order are given by:
\be
E_n^{(0)}=-\frac {m(C_F \alpha_s(\mu))^2}{4n^2},\qquad 
|\Psi^{(0)}_n|^2 = \frac {(m C_F \alpha _s (\mu))^3}{8\pi n^3}.
\ee

First we present an expression for the energy levels valid
up to (relative) order ${\cal O}(\alpha _s^2)$: 
\begin{eqnarray}
E_n &=& E_n^{(0)} \Bigg \{ 1 +\frac {\alpha_s}{\pi} \left (\beta_0+
\frac {\alpha_s}{4\pi} \left (\beta_1+2\beta_0 a_1 \right ) \right )
\left ( L(\mu _a)+\psi_{n+1} \right )
\nonumber \\
&&+ 
\lb \frac{ \alpha_s \beta_0 }{ 2\pi } \rb^2
                      2\left[ (L(\mu_1)+ \psi_{n+1})^2  - \psi_{n+1}'
                      - 2\frac{ \psi_{n+1} + \gamma_E }{ n }
                      + \frac{ \pi^2 }{ 3 } \right]
\nonumber \\
&&+
  \left (\frac {\alpha _s \beta _0}{2\pi} \right )^2
  \Bigg [ \left ( L(\mu_b)+\psi _{n+1} -1 \right )^2
  -1-2 \psi_n^{'} -n \psi_n^{''}+
 \frac {2}{n}
   \left ( \psi _{n+1} +\gamma_E \right ) \Bigg ]
\nonumber \\
&&
-\frac{(C_F \alpha _s)^2}{n}\left ( \frac {11}{16n}-\frac {2}{3}
-\frac {C_A}{C_F}\right )
\Bigg \},
\end{eqnarray}
where $\mu = \mu_{\rm soft}$, 
$$
L(\mu) = \ln \left ( \frac {\mu n}{C_F \alpha _s m} \right )
$$
and the scales $\mu_1,\mu_a,\mu_b$ are defined in Eq.(\ref {scales}).
This result is in agreement with the one obtained in \cite{PY}.

Using the same notations, we obtain the result for the square of the 
wave functions at the origin:
\ba
|\Psi_n|^2 &=& |\Psi_n^{(0)}|^2\; 
{\cal K}(\mu_{\rm hard},\mu_{\rm fact}) \;\Bigg \{ 1
 +\left (\frac {\alpha _s}{2 \pi} \right )
\left (\beta_0+
\frac {\alpha_s}{4\pi} \left (\beta_1+2\beta_0 a_1 \right ) \right )
[ 3 \left (L(\mu_a)+ \psi_{n+1} \right ) -
\nonumber \\
&&  2 \left ( n\psi_n'+\psi_n + \gamma_E \right ) - 1 ]
 + 
\frac{ 3 }{ 2 }
 \lb \frac{ \alpha_s \beta_0 }{ 2\pi } \rb^2
                      2\left[ (L(\mu_1)+ \psi_{n+1})^2  - \psi_{n+1}'
                      - 2\frac{ \psi_{n+1} + \gamma_E }{ n }
                      + \frac{ \pi^2 }{ 3 } \right]
\nonumber \\
&&-2 \lb \frac{ \alpha_s \beta_0 }{ 2\pi } \rb^2 \Bigg [
            L(\mu_1) (2n\psi_{n}'+ 2\psi_{n} + 2\gamma_E + 1) +
2n\psi_n'\left (\psi_n-\gamma_E \right )
+ \psi_n \left (1-2\gamma_E-\frac {2}{n} \right )
\nonumber \\
&&
+2nS_2(n+1)-2n\zeta_3+2\zeta_2 (1+\gamma_E n)+\frac {1}{n}-2\gamma_E^2
\Bigg ]
\nonumber \\
&& +
\left (\frac {\alpha_s
\beta _0}{2\pi} \right )^2
\Bigg [ 3 L(\mu_b)^2-\frac{\left (4 n^2\psi_n'  -2 n\psi_n
+n(5+4\gamma_E)-6 \right ) L(\mu_b)}{n}
\nonumber \\
&&
+
\frac {n^2 \psi_n'''}{6} + \frac{n \psi_n''(1+4 \gamma_E n)}{2}
+
\psi_n'(n^2 \psi_n'+2n(1+3\gamma_E)-5)-
\nonumber \\
&&
\frac {\psi_n (n \psi_n + 3(n - 1)+4 \gamma_E n)}{n}
+
\frac {(6-5 n+n^2+2 n^3 \zeta _3 - n^4 \zeta _4)}{n^2}+
\nonumber \\
&&
\frac {2 \gamma_E (2 n^3 \zeta _3 -n^2 \zeta _2 
-\frac {3}{2} +n)}{n}+4n^2S_3(n)-2nS_2(n) \Bigg ] +
\nonumber \\
&&
(C_F \alpha _s)^2
\left [-\frac {37}{24n^2}-\left (\frac {2}{3} + \frac {C_A}{C_F}
\right )\left (\ln \left ( \frac {C_F \alpha _s m}{2n \mu_{\rm fact}} \right )
+\psi_{n}+ \gamma_E - \frac {1}{n} \right ) \right ]
\Bigg \}.
\ea
In the above expression, ${\cal K}(\mu_{\rm hard},\mu_{\rm fact})$ stands
for the NNLO hard renormalization factor given explicitly 
in Eq.(\ref {hard}).

\section {Theoretical 
moments and numerical analysis of sum rules}

The theoretical moments can be conveniently separated
into the contributions of the perturbative resonances
and of the perturbative continuum:
\be
{\cal M}_n = {\cal P}_n + {\cal C}_n.
\ee

The resonance contribution reads:
\be
{\cal P}_n = 
6N_c\pi^2 \left(\frac {M_1}{2m} \right )^{2n}
\sum \limits_{k=1}^{\infty} 
\frac {|\Psi_k|^2}{m^3 \left (1+\frac {E_k}{2m} \right )^{2n+1}},
\label {pn}
\ee
and the continuum contribution is defined as the integral 
of the function $R(s)/s^{n+1}$ over $s$ (see Eq.(\ref{rs}))
taken above the threshold.

We stress that the above expression for the moments differs
from the result one gets, merely integrating the corrections to the 
Green function. 
The difference is due to the fact, that in Eq.(\ref {pn}),
we calculate the corrections separately to the numerator
and the denominator. 
Working in the limit $\sqrt{n}\alpha _s \sim 1$, it is possible to
expand the denominators in Eq. (\ref {pn}) around their values
for the exact Coulomb problem; this would produce (parametrically) an  
${\cal O}(\alpha_s)$  corrections 
to the moments. However, there is 
a serious numerical difference between the expanded and not expanded 
denominators. The origin of the problem is related to the 
large values of $\beta_0$ and $a_{1,2}$ entering the
potential. This effectively translates into the large values of
the corrections to the energy levels; 
for this reason, the expansion of the 
denominators is not justified in our opinion.

In contrast to the contribution 
of the perturbative resonances to the moments of the spectral function,
the contribution of the continuous
spectrum to the moments behaves nicely, as far as its perturbative
expansion is concerned. The contribution of the perturbative
continuum to the theoretical moments is obtained by numerical 
integration.

For numerical analysis of the
sum rules, we  use the value
of the strong coupling constant $\alpha _s(M_Z)$ equal
to the world average value $\alpha_s(M_Z)=0.118$. This value
of the strong coupling constant is evolved down to a 
required scale $\mu$ using the two--loop renormalization group
evolution equation.  We will later comment on the sensitivity
of our results to the value of the strong coupling constant 
at the $Z$--resonance.

We also parameterize the unknown contribution of the experimental
continuum $R_c(s)$ in Eq.(\ref {momentexp}) by a constant, which 
we vary between $0$ and $2$. The value of $s_0$ in Eq.(\ref {momentexp})
equals to $(2\times 5.927\;\mbox{GeV})^2$, i.e. the continuum contribution 
starts
at the threshold of the open $B\bar B$ production. To suppress
the influence of this unknown contribution we have to go to rather
high values of $n$. We have  chosen $n$  equal to $14,16,18$,
for our analysis.

We fix the values of the hard and factorization 
scales at $5$ GeV and examine the value of the 
$b$-quark mass as a function of the soft scale $\mu_{\rm soft}$.
Our results are presented in Table 1. The values of the kinetic mass at 
$\mu_Q=1$ GeV are found using Eq.(\ref{polemass}) treated with the
necessary accuracy.

\begin{table}
\begin{center}
$$
\begin{array}{||c||c||c|c||c|c||c|c||c||}
\hline
&
&
\multicolumn{2}{|c|}{14}&
\multicolumn{2}{|c|}{16}&
\multicolumn{2}{|c|}{18}& {\rm PT~for}~{\cal M}_{14}\\
\cline{3-8}
{\rm Scales}& {\rm Order~of~PT}
&m_{\rm pole} & m_{\rm kin} &  m_{\rm pole} &  m_{\rm kin}  
& m_{\rm pole} & m_{\rm kin} & {\rm for~NNLO}~m_{\rm kin} 
\\ \hline \hline
\mu_{\rm soft} = 4.5 ~{\rm GeV} &{\rm LO}&  4.69 & 4.57 & 4.7 
&  4.57&  4.7&  4.575& 0.97\\ 
\cline{2-9}
\mu_{\rm hard} = 5 ~{\rm GeV} &{\rm NLO}
& 4.74&  4.49&  4.75 &  4.50 & 4.76
& 4.51 
& 0.61\\ \cline{2-9}
\mu_{\rm fact} = 5 ~{\rm GeV}
&{\rm NNLO} &  4.89 &  4.51 & 4.90 
&  4.52 &  4.91 &  4.53  & 0.69\\ \hline \hline
\mu_{\rm soft} = 3.5 ~{\rm GeV} &{\rm LO} & 4.73 & 4.59 
& 4.73& 4.59& 4.73 
& 4.595 &0.95\\ \cline{2-9}
\mu_{\rm hard} = 5 ~{\rm GeV} &{\rm NLO} & 4.77& 4.50
& 4.78& 4.511
& 4.79& 4.52 &0.55\\ \cline{2-9}
\mu_{\rm fact} = 5 ~{\rm GeV}
&{\rm NNLO}& 4.95& 4.54& 4.96 & 4.55 & 4.96&
4.55& 0.69\\
\hline \hline 
\mu_{\rm soft} = 2.5 ~{\rm GeV}
&{\rm LO} & 4.79& 4.63&
4.79& 4.63 & 4.79& 4.63&0.87\\ \cline{2-9} 
\mu_{\rm hard} =5 ~{\rm GeV} &{\rm NLO} & 4.82& 4.505& 4.83&
4.517 & 4.84&  
4.525& 0.40\\ \cline{2-9} 
\mu_{\rm fact} = 5 ~{\rm GeV}&{\rm NNLO}& 5.08 &
4.595 &  5.08& 4.595 &  5.09 & 4.6 &0.69\\ 
\hline \hline
\mu_{\rm soft} = 2 ~{\rm GeV} &{\rm LO} & 4.84 & 4.67 
& 4.84& 4.666& 4.84 
& 4.66 &0.72\\ \cline{2-9}
\mu_{\rm hard} = 5 ~{\rm GeV} &{\rm NLO}& 4.84& 4.49
& 4.85& 4.505
& 4.86& 4.514 &0.24\\ \cline{2-9}
\mu_{\rm fact} = 5 ~{\rm GeV}
&{\rm NNLO}& 5.21& 4.66& 5.21 & 4.66 & 5.21&
4.66&0.69 \\
\hline
\end{array}
$$
\vspace*{0.5cm}
\caption{The kinetic mass $m_{\rm kin}$ for $\mu_Q = 1$ GeV 
as a function of the soft renormalization scale for the moments 
$n=14,16,18$ and successive approximations for 
${\cal M}_{14}$ evaluated for NNLO $m_{\rm kin}$.} 
\end{center}
\end{table}

 The last column in Table 1 demonstrates how the perturbation theory 
works for the theoretical moment ${\cal M}_{14}$ at LO, NLO and NNLO if
the NNLO value of the kinetic mass is used as an input. For comparison
we quote here also the values of the ${\cal M}_{14}$ moment for the 
pole mass, which corresponds to the formal limit $\mu _Q \to 0$ in our
approach. For $\mu_{\rm soft}=3.5$ GeV the NNLO pole mass equals
to $4.95$ GeV. Calculating ${\cal M}_{14}$ with this mass we obtain: 
${\cal M}_{14}^{\rm LO} = 0.2$, ${\cal M}_{14}^{\rm NLO} = 0.25$ and
${\cal M}_{14}^{\rm NNLO} = 0.69$.

 From Table $1$ one can see two things -- the perturbation 
theory for the pole
mass behaves in a way, that  does not show 
any sign of convergence; the NNLO corrections 
normally exceed the NLO ones.
Moreover, the pole mass strongly
depends on the soft renormalization scale.
This picture is consistent 
with the expectation of the irreducible ambiguity of order
$\Lambda _{\rm QCD}$ in the pole mass.

For the low-scale mass the situation is different
in both respects. The first terms of the perturbation 
series  are sign alternating (if taken seriously, this  feature
signals that we are on the right way).
Also, the low-scale running 
mass exhibits  only a moderate dependence on $\mu_{\rm soft}$
in a relatively wide range of the soft renormalization scale. 
The width of this range   depends on the
order of perturbation theory we consider; it also  depends
on the initial value of $\alpha _s (M_z)$. 
The most stable picture emerges at the NLO,
while the inclusion of the NNLO effects makes the result less stable. 
This partial loss of stability is the consequence of the very
large value of the second order correction $a_2$ to the perturbative 
quark antiquark potential. 

If the sign alternating behavior of the perturbation series 
for $m_{\rm kin}$ is taken seriously, one can try to perform
some transformation of the perturbation series to accelerate
the convergence. One of numerous options
is the Euler transformation.
It is interesting to observe that this transformation 
indeed brings the values of $m_{\rm kin}$ obtained 
for various values of $\mu_{\rm soft}$ closer to each other.

The Euler transformation works in the following way. Imagine
we have a series $f(z) = \sum (-1)^n c_n z^n$. Then the faster
convergent  approximation for $f(z)$ is given by:
\be
f(z) = \frac {1}{1+z} \left ( c_0 -(c_1-c_0) \left ( \frac {z}{1+z}
\right ) +(c_2-2c_1+c_0)  \left ( \frac {z}{1+z} \right )^2+...
\right ).
\label {ET}
\ee
We restricted our consideration to two orders of perturbation theory
which we can use to determine the expression for the mass.

We then identify $z$ in the previous formula with $\alpha(\mu_{\rm
soft})$. Then we use the results in  Table 1 for $n=16$ and 
$\mu_{\rm soft} = 4.5,\;2.5,\;2$ GeV. 
For the sake of illustration,  we present such 
calculation for $\mu_{\rm soft}=4.5$ GeV:
\be
\tilde m =4.57-0.07+0.02 = 4.57\left (1-0.07 \alpha_s+0.09 \alpha _s^2
\right ),\qquad \alpha_s = \alpha _s(4.5) = 0.22.
\ee
We then use Eq.(\ref {ET}) and the above formula 
to obtain a faster convergent series:
\be
m_{\rm kin} = \frac {4.57}{1+\alpha _s} \left (1
+0.93\frac {\alpha _s}{1+\alpha_s}+0.95
\frac {\alpha _s ^2}{(1+\alpha_s)^2}+... \right ) = 4.49~{\rm GeV}.
\ee
The result does not change notably, 
as compared to the ``naive'' summation of the
$\alpha _s$ series, indicating that in this case the numerical value
of $\alpha _s$ is fairly small.

In the same way, we obtain the new values of the kinetic masses for
the $\mu_{\rm soft} = 2.5$ and $2$ GeV. The results are:
for $\mu_{\rm soft} = 2.5$ GeV, we obtain $m_{\rm kin}=4.52$ GeV and
for  $\mu_{\rm soft} = 2$ GeV, $m_{\rm kin}=4.53$ GeV. 

Clearly, such transformation  cannot be rigorously justified; 
however, the fact that the numbers come closer to each other 
looks gratifying. Moreover, both numbers appear to become closer to the 
value of the NNLO mass, which can be obtained by examining the
region of relative stability $\mu_{\rm soft} > 2.5-3 $ GeV, which
we use below to estimate the value of the $b$ quark mass.

Let us also comment on how the choice of the value of 
$\alpha _s (M_Z)$ is reflected on our result. The important
point is that this dependence is rather moderate, since the change
in the initial value of $\alpha _s (M_Z)$ is roughly equivalent
to the change in $\mu_{\rm soft}$. As we always work in the 
region where the dependence of $m_{\rm kin}$ on the 
normalization scale of the coupling constant is relatively 
weak, the same  equally applies 
to the dependence on the  initial value of  $\alpha _s$ at 
the $Z$--resonance. The dependence
on the factorization scale is much weaker than on the soft
renormalization scale and amounts to a variation 
of at most $20$ MeV in value of the kinetic mass.

We arrive finally at the following estimate for the kinetic 
mass  extracted from the QCD sum rules:
\be
m_{\rm kin} (1{\rm GeV}) = 4.56 \pm 0.06~{\rm GeV}.
\ee

We stress, that the error in the above estimate 
is  primarily of the theoretical origin. The 
experimental errors in the masses and electronic decay widths 
of the $\Upsilon$  resonances, 
as well as a poor knowledge of the continuum part of the 
observable spectrum, are minor effects as compared with, e.g., 
the soft scale dependence of the kinetic mass.
For the above estimate, we use our results for $\mu_{\rm soft}$ 
between 4.5 and 2.5 GeV. For lower scales, the perturbation 
series for the moments do not look reliable enough. 

After the value of the pole mass is found, one can obtain 
the estimate of the $\overline {MS}$ mass $\bar m$. 
To order ${\cal O}(\alpha _s ^2)$ the corresponding 
equation was given in \cite{CzMeUr}. One obtains:
\be
\bar m(\bar m) = 4.20 \pm 0.1~{\rm GeV}. 
\ee

Using Eq. (\ref{polemass}), one can derive 
the evolution equation with respect
to $\mu_Q$ and calculate the kinetic mass at different 
normalization scales. 
The only thing to be remembered is that a choice of the 
normalization scale $\mu_Q$ is limited by two inequalities: 
$\Lambda_{\rm QCD} \ll \mu_Q \ll m $. From this point of view, the
pole mass, which formally corresponds to the limit $\mu_Q \to 0$,
is seen to be a completely artificial notion, since it includes
a nonperturbative contribution treated in terms of the perturbation
theory.

Our NNLO results may look incomplete since we use  relation 
(\ref{polemass}) between the pole and kinetic masses, which is valid 
to $\mu_Q\alpha_s^2$ order, while formally one needs to know this 
relation more accurately, to  $\mu_Q\alpha_s^3$ order. 
However, a careful examination based on the BLM--estimate 
(\ref {BLM})
of the ${\cal O}(\mu_Q\alpha_s^3)$ terms, 
shows that  unknown  corrections in that order in the relation
between the pole mass and the kinetic mass
cannot drastically change our results for the kinetic mass 
in the NNLO approximation.

In any case, working 
with the values of the $\mu_{\rm soft}$ where perturbation 
theory seems to be reliable, we think we can provide
a reasonable estimate for the value of the low-scale running
mass to NNLO. 

\section{Conclusions}

In this paper, we have determined the 
$b$ quark low-scale running mass \cite{5infty}
from the analysis of the QCD sum rules in the next-to-next-to-leading order.
We have shown that the use of this mass significantly
improves the convergence of the 
perturbation series for the moments of the 
spectral density. As the result of our analysis we obtain 
the value of the kinetic mass normalized at $1$ GeV:
$m_{\rm kin}(1{\rm GeV}) = 4.56 \pm 0.06~{\rm GeV}$ and the 
corresponding value of the  $\overline{MS}$
mass $\bar m$: $\bar m(\bar m) = 4.20 \pm 0.1~{\rm GeV}$.

In our opinion, the pole mass of the $b$ quark cannot be  
reliably determined from the sum rule analysis.
We have shown, that the NNLO order corrections to the 
pole mass are typically larger than the NLO ones.
Also the value of the pole mass of the $b$ quark
is very sensitive to the scale of the strong coupling
constant, that is used in the analysis. We think that 
these features  are in accord with the fact  that the pole 
mass of the quark cannot be reliably defined theoretically 
and suffers from an irreducible ambiguity of the order of 
$\Lambda_{\rm QCD}$ \cite{Bigi,BeBr}. 

As a byproduct of our study 
we have obtained  the NNLO analytical expression for the cross section  
$e^+e^- \to Q \bar Q$ of the quark antiquark pair production 
in the threshold region. We have also given the NNLO expressions
for the energy levels and the wave functions at the origin 
for the $^1S_3$ bound states of $Q\bar Q$.

\section{Acknowledgments}

We are grateful to V.L. Chernyak, A.A. Penin and N.G. Uraltsev for useful
discussions.  We would like to thank S.I. Eidelman for the discussion
of the experimental data on $R(s)$.
The research of K. M. was supported in part by BMBF
under grant number BMBF-057KA92P, and by Graduiertenkolleg
``Teilchenphysik'' at the University of Karlsruhe.
A.Y. was partially supported by the Russian
Foundation for Basic Research under grant number 98-02-17913.


\begin{thebibliography}{10}

\bibitem{Vol1} V.A. Novikov {\it et al}, Phys. Rev. Lett. {\bf 38} (1977), 626;\\
 V.A. Novikov {\it et al}, Phys. Rep. {\bf C41} (1978), 1.

\bibitem{Vol2} M.B. Voloshin, ITEP preprint 1980-21, unpublished;\\
M.B. Voloshin and Yu.M. Zaitsev, 
Sov. Phys. Usp. {\bf 30} (1987), 553.

\bibitem{Vol} M.B. Voloshin, Int. Journ. Mod. Phys. {\bf A10} (1995), 2865.

\bibitem{JP} M. Jamin and A. Pich, Nucl. Phys. {\bf B507} (1997), 334. 

\bibitem{KPP} J.H. K\"uhn, A.A. Penin and A.A. Pivovarov,  hep-ph/9801356. 

\bibitem{Penin} A.A. Penin and A.A. Pivovarov, hep-ph/9803363.

\bibitem{Hoangb} A.H. Hoang, hep-ph/9803454.

\bibitem{vosem} I.I. Bigi and N.G. Uraltsev, 
Phys. Lett. {\bf B321} (1994), 412.

\bibitem{Bigi} I. Bigi, M. Shifman, N. Uraltsev and A. Vainshtein,
Phys. Rev. {\bf D50} (1994), 2234.

\bibitem{BeBr} M. Beneke and V. Braun, Nucl. Phys. {\bf B246} (1994), 301.

\bibitem{Klecture} N.G. Uraltsev,  
Lectures given at ``Heavy Flavour Physics: A Probe of Nature's
Grand Design'' (International School of Physics ``Enrico Fermi"),
Varenna, July 7-18 1997. To appear in the Proceedings, IOS Press,
Amsterdam,  hep-ph/9804275.

\bibitem{5infty}  I. Bigi, M. Shifman, N. Uraltsev and A. Vainshtein,
Phys. Rev. {\bf D56} (1997), 4017.

\bibitem{Hoang} A.H. Hoang, Phys. Rev. {\bf D56} (1997), 5851.

\bibitem{HT} A.H.Hoang and T. Teubner, preprint UCSD/PTH 98-01,
DESY 98-008, hep-ph/9801397.

\bibitem{top} K. Melnikov and A. Yelkhovsky, hep-ph/9802379, Nucl.
Phys. {\bf B}, in press.

\bibitem{CzMeUr} A. Czarnecki, K. Melnikov and N. Uraltsev,
Phys. Rev. Lett. {\bf 80} (1998), 3189.

\bibitem{BLM} S.J. Brodsky, G. Lepage and P. Mackenzie,
Phys. Rev. {\bf D28} (1983), 28.


\bibitem{LL}
L.D. Landau and  E.M. Lifschitz, 
{\it Relativistic Quantum Theory}, Part 1 
(Pergamon, Oxford, 1974).

\bibitem{NABreit} S.N. Gupta and S.F.Randford, Phys. Rev. {\bf D24}
(1981), 2309, (E) {\it ibid} {\bf D25} (1982), 3430;\\
S.N. Gupta, S.F.Randford and W.W. Repko, Phys. Rev. {\bf D26} (1982), 3305.

\bibitem{Fishler} W. Fischler, Nucl. Phys. {\bf B129} (1977), 157;\\
A. Billoire, Phys. Lett. {\bf B92} (1980), 343.


\bibitem{Peter} M. Peter, Nucl. Phys. {\bf B501} (1997), 471.

\bibitem{CzM} A. Czarnecki and K.Melnikov, 
Phys.\ Rev.\ Lett. {\bf 80} (1998), 2531.

\bibitem{MS} A.I. Milshtein and V.M. Strakhovenko,
Phys. Lett. {\bf 90A} (1982), 447.


\bibitem{PY} A. Pineda and F.J. Yundurain, hep-ph/9711287.

\end{thebibliography}

\end{document}